%% file: main.tex
\documentclass[aps,prl,superscriptaddress,showkeys,showpacs,twocolumn,longbibliography,nofootinbib]{revtex4-1}
\usepackage[utf8]{inputenc}
\usepackage{amsmath}
\usepackage{amsfonts}
\usepackage{amsthm}
\usepackage{amssymb}
\usepackage{float}
\usepackage{braket}
\usepackage{graphicx}
\usepackage{url}
\usepackage[colorlinks=true, pdfstartview=FitV, linkcolor=red, citecolor=blue, urlcolor=blue]{hyperref}
\usepackage{slashed}
\usepackage[normalem]{ulem}

\graphicspath{{./figures/}}

\newcommand{\be}{\begin{equation}}
\newcommand{\ee}{\end{equation}}
\newcommand{\bea}{\begin{eqnarray}}
\newcommand{\eea}{\end{eqnarray}}

\newcommand{\Tr}{\mathrm{Tr}}

\newcommand{\beq}{\begin{equation}}
\newcommand{\eeq}{\end{equation}}

\newcommand{\Ldyn}[1]{L_{\mathrm{dyn},#1}}
\newcommand{\Lext}[1]{L_{\mathrm{ext},#1}}
\newcommand{\jext}{j_{\mathrm{ext}}}
\newcommand{\Hext}{H_{\mathrm{ext}}}

\begin{document}
\title{Real-time non-perturbative dynamics of jet production in Schwinger model:\\ 
quantum entanglement and vacuum modification}

\author{Adrien Florio}
\email[]{aflorio@bnl.gov}
\affiliation{Department of Physics, Brookhaven National Laboratory, Upton, New York 11973-5000, USA}

\author{David Frenklakh}
\email[]{david.frenklakh@stonybrook.edu}
\affiliation{Center for Nuclear Theory, Department of Physics and Astronomy, Stony Brook University, Stony Brook, New York 11794-3800, USA}

\author{Kazuki Ikeda}
\email[]{kazuki.ikeda@stonybrook.edu}
\affiliation{Center for Nuclear Theory, Department of Physics and Astronomy, Stony Brook University, Stony Brook, New York 11794-3800, USA}
\affiliation{Co-design Center for Quantum Advantage, Department of Physics and Astronomy, Stony Brook University, Stony Brook, New York 11794-3800, USA}

\author{\mbox{Dmitri Kharzeev}}
\email[]{dmitri.kharzeev@stonybrook.edu}
\affiliation{Department of Physics, Brookhaven National Laboratory, Upton, New York 11973-5000, USA}
\affiliation{Center for Nuclear Theory, Department of Physics and Astronomy, Stony Brook University, Stony Brook, New York 11794-3800, USA}
\affiliation{Co-design Center for Quantum Advantage, Department of Physics and Astronomy, Stony Brook University, Stony Brook, New York 11794-3800, USA}

\author{Vladimir Korepin}
\email[]{vladimir.korepin@stonybrook.edu}
\affiliation{C.N. Yang Institute for Theoretical Physics, Stony Brook University, Stony Brook, New York, 11794-3840, USA}

\author{Shuzhe Shi}
\email[]{shuzhe-shi@tsinghua.edu.cn}
\affiliation{Department of Physics, Tsinghua University, Beijing 100084, China}
\affiliation{Center for Nuclear Theory, Department of Physics and Astronomy, Stony Brook University, Stony Brook, New York 11794-3800, USA}

\author{Kwangmin Yu}
\email[]{kyu@bnl.gov}
\affiliation{Computational Science Initiative, Brookhaven National Laboratory, Upton, New York 11973-5000, USA}

\bibliographystyle{unsrt}

\begin{abstract}
The production of jets should allow testing the real-time response of the QCD vacuum disturbed by the propagation of high-momentum color charges.
Addressing this problem theoretically requires a real-time, non-perturbative method.
It is well known that the Schwinger model [QED in $(1+1)$ dimensions] shares many common properties with QCD, including confinement, chiral symmetry breaking, and the existence of vacuum fermion condensate. 
As a step in developing such an approach, we report here on fully quantum simulations of a massive Schwinger model coupled to external sources representing quark and antiquark jets as produced in $e^+e^-$ annihilation.
We study, for the first time, the modification of the vacuum chiral condensate by the propagating jets and the quantum entanglement between the fragmenting jets. Our results indicate strong entanglement between the fragmentation products of the two jets at rapidity separations $\Delta \eta \leq 2$, which can potentially exist also in QCD and can be studied in experiments.
\end{abstract}

\maketitle

\textit{Introduction:}
The discovery of jets played a crucial role in establishing Quantum Chromodynamics (QCD) as the theory of strong interactions, see~\cite{Sterman:2004pd, Dokshitzer:2000yu} for reviews. The production of the initial high momentum partons is a short-distance process that can be described in perturbative QCD due to asymptotic freedom. However, as the initial partons keep radiating gluons and quark-antiquark pairs as described by QCD evolution equations, the characteristic virtuality decreases, and non-perturbative phenomena should come into play. 

In particular, one expects that the propagating color charges will disturb the non-perturbative QCD vacuum, and the corresponding real-time response should contain valuable information about the vacuum structure. Moreover, the initial partons should be entangled by the production process, but whether any trace of this entanglement can be found in fragmenting jets is not clear. The answers to these questions lie outside of the realm of perturbative QCD, and finding them requires a real-time, non-perturbative method. 

Such an approach is enabled by the advent of quantum simulations. Unfortunately, the case of real $(3+1)$ dimensional QCD is still out of reach for the existing quantum hardware, as well as for real-time simulations on classical computers. However one can start developing real-time non-perturbative methods using simpler models in lower number of space-time dimensions. 

In this respect QED in $(1+1)$ dimensions (the Schwinger model~\cite{Schwinger:1962tp}) holds a special place: just like QCD, it possesses confinement, chiral symmetry breaking, and fermion condensate \cite{Coleman:1975pw}. In the massless fermion limit, the theory is exactly solvable by bosonization, and admits a dual description in terms of a free massive scalar theory. In 1974, Casher, Kogut, and Susskind~\cite{Casher:1974vf} proposed to model quark-antiquark production in $e^+e^-$ annihilation by coupling Schwinger model to external sources propagating along the light cone. 

An explicit analytical solution of this model has been found in~\cite{Loshaj:2011jx, Kharzeev:2012re}, where this setup was also used to describe jet quenching in heavy ion collisions by introducing in-medium scattering of the sources, and the anomalous enhancement of soft photon production in jet fragmentation~\cite{Kharzeev:2013wra} observed by the DELPHI Collaboration~\cite{DELPHI:2010cit}. 

A more realistic extension of this approach is based on a massive Schwinger model, which in the bosonized description is dual to an interacting meson theory. In this case, the model is no longer analytically solvable, and so a numerical approach is necessary. The first study 
of this setup was carried out in~\cite{Hebenstreit:2014rha} using a numerical classical-statistical approach. Coupling the Schwinger model to an external Yukawa theory has also been used to mimic the propagation of jets through a thermal environment~\cite{deJong:2021wsd}. Various other aspects of the Schwinger model have also been addressed using quantum simulations, see~\cite{Klco:2018kyo, Butt:2019uul, Magnifico:2019kyj, Shaw:2020udc, Kharzeev:2020kgc, Ikeda:2020agk, Rigobello:2021fxw,Ikeda:2023zil,Ikeda:2023vfk} for examples and~\cite{Bauer:2022hpo} for a recent review of quantum simulations. 

In this work, using the massive Schwinger model coupled to external sources, we perform the first fully quantum simulation of jet production. In particular, we focus on real-time, non-perturbative effects that have not been studied before: the modification of the vacuum structure and the entanglement between the produced jets.

\textit{The model:}
We use the massive Schwinger model Hamiltonian in temporal gauge $A_0=0$ in the presence of an external current $\jext^{\mu}$ describing the produced jets:
\begin{align}
H^C &= H^C_S + \Hext^C\ ,\label{eq:Ham}\\
H_S^C &=\int dx \left [ \frac{1}{2}E^2+\bar{\psi}(-i\gamma^1\partial_1+g\gamma^1A_1+m)\psi \right ], \\
\Hext^C &= \int dx\, \jext^1 A_1 \ ,
\end{align}
where $A_\mu$ is the $U(1)$ gauge potential, $E=-\dot{A^1}$ is the corresponding electric field, $\psi$ is a two-component fermionic field, $m$ is the fermion mass, and $\gamma^\mu$ are two-dimensional $\gamma$-matrices satisfying Clifford algebra; we use $\eta_{\mu\nu}=\mathrm{diag}(1,-1)$ as our metric. The superscript $C$ stands for ``continuum".

The effect on the theory of the interaction with the external source $\Hext$ is to modify Gauss law to  
\begin{equation}
\partial_1 E - j^0 = \jext^0 \ .
\end{equation}
with $j^0 = g\,\bar\psi\gamma^0\psi$.
In other words, the theory is gauge invariant up to the presence of the external charge $\jext^0$; the external current is a ``defect" of the $U(1)$ gauge transformation.

 To mimic production of a pair of jets in $e^+e^-$ annihilation, we choose the external current to represent  charges of opposite sign flying apart along the light cone:
\begin{align}
\begin{aligned}
    \jext^0(x,t)=\;&g[\delta(\Delta x - \Delta t) - \delta(\Delta x + \Delta t)]\theta(\Delta t)\,, \label{eq:jext}\\
    \jext^1(x,t)=\;&g[\delta(\Delta x - \Delta t) + \delta(\Delta x + \Delta t)]\theta(\Delta t) \ ,
\end{aligned}
\end{align}
where ($t_0$, $x_0$) is the time and position of a point where the jet pair is produced, and $\Delta x \equiv x - x_0$ and $\Delta t \equiv t - t_0$ are the space and time distance from this position.

Note that in principle one could replace the external probe charges by ``hard" dynamical fermions,  which can, for instance, be produced by short lived pulses of electric fields. This has been done in~\cite{Hebenstreit:2014rha}, where it was found that, at least within the semiclassics, the use of external charges is a very good approximation to a pair of dynamical relativistic ``hard" fermions. This motivates us to restrict ourselves to the simpler case of external currents.

Our goal is to study the modification of the vacuum due to the presence of the external sources~\eqref{eq:jext}. To this end, we evolve the ground state of the massive Schwinger model with the time-dependent Hamiltonian~\eqref{eq:Ham}. In order to solve this problem, we need to discretize space-time and approximate the theory by a finite-dimensional Hilbert space.

\textit{Lattice model:} We begin by discretizing space in a lattice of $N$ points with lattice spacing $a$. We choose to work with staggered fermions $\chi_n$~\cite{Kogut:1974ag, Susskind:1976jm}. We use a non-compact formulation for the $U(1)$ gauge fields, and introduce a lattice electric field operator $L_n= E(an)/g$, a lattice vector potential $\phi_n=ag\,A_1(an)$, and a link operator $U_n = e^{-iagA_1(an)}$. We further impose open-boundary conditions (OBC) $\chi_{N+1}=L_{N}=0$ on the fermion and gauge fields. Using the Dirac matrices $\gamma^0 = \sigma_z \equiv Z$, $\gamma^1 = i\,\sigma_y \equiv i\,Y$, the Hamiltonian is
\begin{align}
H^L(t) &= H^L_S + \Hext^L(t) \ ,\label{eq:Haml}\\
H_S^L &= -\frac{i}{2a}\sum_{n=1}^{N-1}
\big[U_n^\dag\chi^\dag_{n}\chi_{n+1}-U_n\chi^\dag_{n+1}\chi_{n}\big]
\nonumber\\
&+\frac{ag^2}{2}\sum_{n=1}^{N-1}L_n^2+ m\sum_{n=1}^{N} (-1)^n \chi^\dag_n\chi_n \ , \\
\Hext^L(t) &= \frac{1}{g}\sum_{n=1}^{N-1} \jext^1(a\,n, t)\phi_n \ ,
\end{align}
where the superscript $L$ stands for ``lattice". Even in the presence of point charges, Gauss law is well defined when integrated over a lattice spacing and reads
\begin{align}
  L_{n} - L_{n-1} - Q_n &= \frac{1}{g}\int_{(n-1/2)a}^{(n+1/2)a} \mathrm{d} x\, \jext^0(x,t) \ , \label{eq:GaussLat}
\end{align}
with $Q_n = \chi_n^\dagger \chi_n - (1-(-1)^n)/2$ the lattice charge density operator. For the rest of this work, we insert the sources at the center of our lattice, $x_0 = a\left(\frac{N+1}{2}\right)$, at time $\frac{t_0}{a} = 1$.

Before proceeding with the time evolution, we take advantage of the fact that the gauge fields are non-dynamical in $(1+1)$ dimensions to express them in terms of fermionic operators through Gauss law. This has the advantage of drastically reducing the size of the discrete Hilbert space needed
down to $2^N$, at the cost of introducing non-localities. The former turns out to outweigh the latter for the method we use (direct diagonalization, or ``exact diagonalization" of the Hamiltonian), see also the Supplementary Material.

\begin{figure*}
    \centering
    \includegraphics[width=0.5\textwidth]{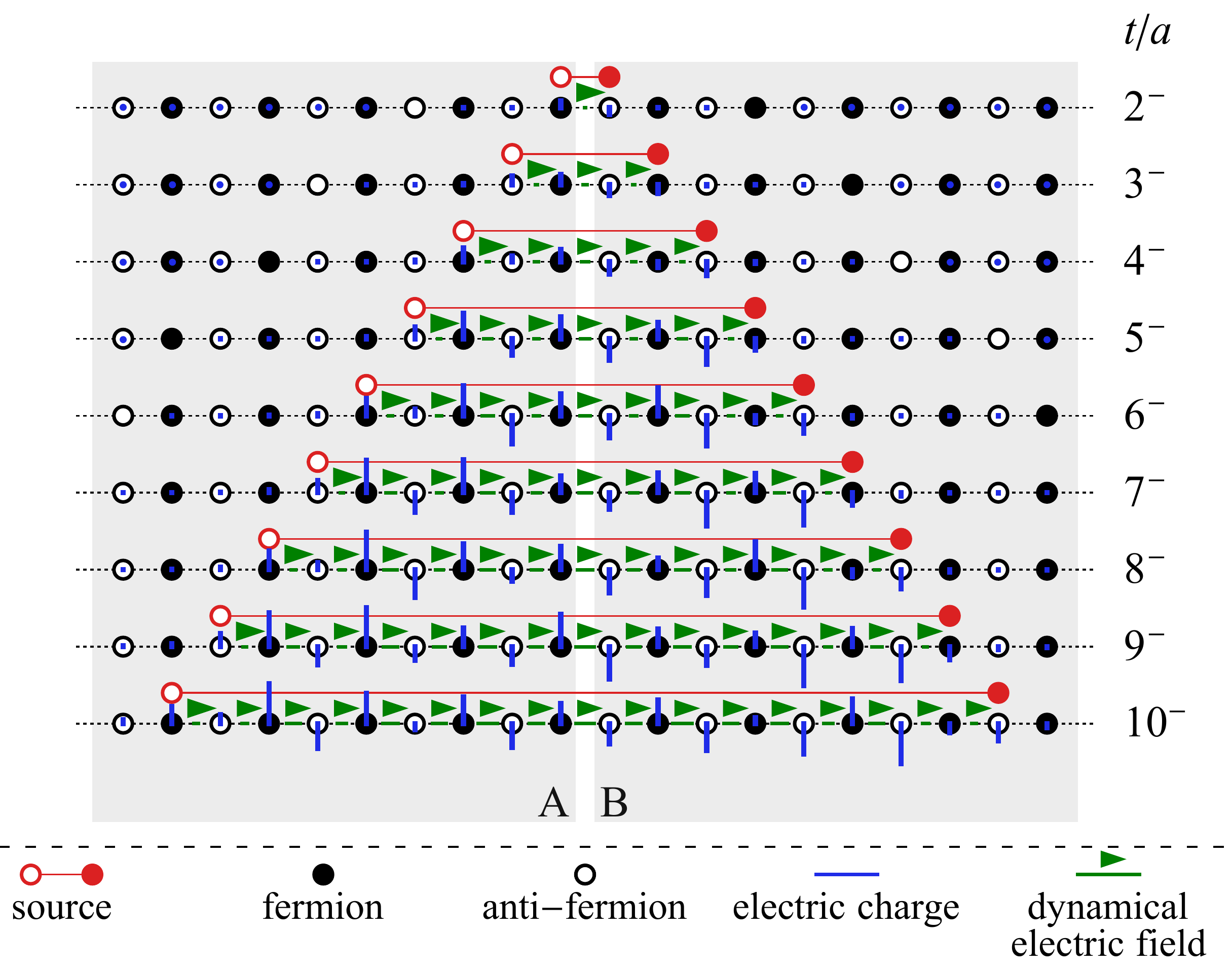}\qquad
    \includegraphics[width=0.3\textwidth]{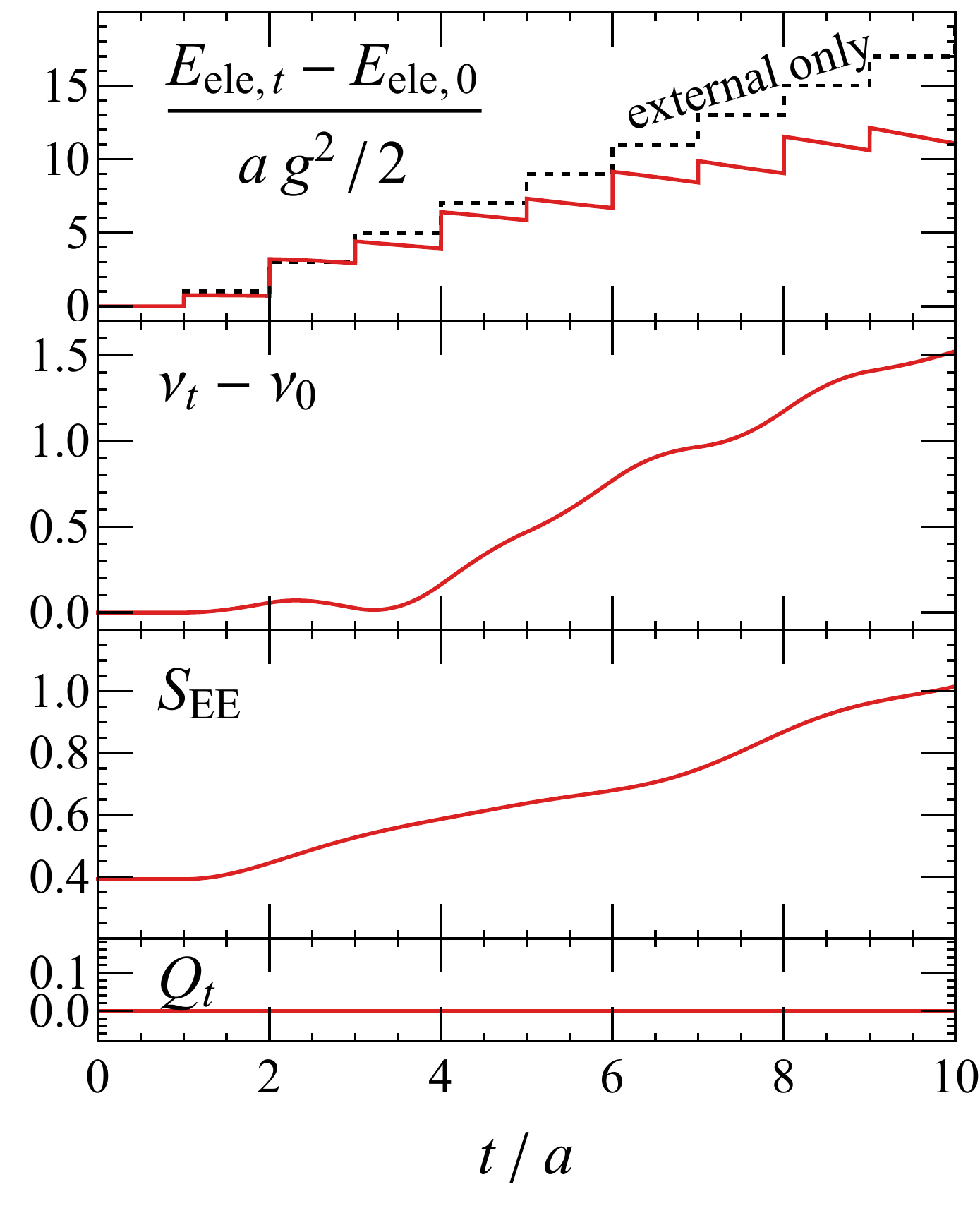}
    \caption{(Left) Time evolution of the local charge density (vertical bars) and of the electric field (arrows), with vacuum expectation values subtracted. Black(white) even(odd)-sites correspond to (anti)fermions. The position of the external sources is shown above each configuration. From top to bottom, the rows are for time values (in units of lattice spacing $a$) $t/a = 2^-$ to $10^-$, where $n^-\equiv n-\varepsilon$ with $\varepsilon$ being an arbitrarily small positive number. 
    (Right) (from top to bottom) Time evolution of electric energy, scalar fermion density, entanglement entropy, and electric charge. 
    Dotted lines in the first panel show the  electric energy generated by the external sources. The value of the vacuum fermion condensate integrated over the lattice length is $\nu_0 = -5.16$.}
    \label{fig:1}
\end{figure*}

We then use the remaining freedom to perform a space-only dependent gauge transformation to set all gauge links to unity. The explicit gauge transformation which achieves this result is $\Omega_1=1, \Omega_n=\prod_{i=1}^{n-1} U_i^\dagger$ \cite{Ikeda:2020agk}. Note that the existence of such a transformation is a peculiarity of $(1+1)$ dimensions and is related to the fact that the gauge field is not dynamical. We then rewrite $L_n=\Ldyn{n} + \Lext{n}$ and solve Gauss law \eqref{eq:GaussLat} as follows:
\begin{align}
  \Ldyn{n} &= \sum_{i=1}^n Q_i\,, \\
  \Lext{n}(t) &= -\theta\left(t-t_0 - \left|x-x_0+\frac{a}{2}\right|\right) \ .
\end{align}
The non-locality is contained in the dynamical gauge field and the external sources create a chain of electric fluxes between them.

The Hamiltonian \eqref{eq:Haml} is now directly suitable for diagonalization. However, having in mind future quantum computing applications, we have used an equivalent form in terms of Pauli matrices $X,Y,Z$, or ``spin" degrees of freedom.  We employ the Jordan--Wigner transformation~\cite{Jordan:1928wi}
\begin{align}
\begin{aligned}
 \chi_n &=\frac{X_n-iY_n}{2}\prod_{j=1}^{n-1}(-i Z_j),\\
 \chi^\dag_n &=\frac{X_n+iY_n}{2}\prod_{j=1}^{n-1}(i Z_j),
\end{aligned}
\label{eq:jordan_wigner}
\end{align}
to obtain
\begin{align}
     H^L(t)=&\frac{1}{4a}\sum_{n=1}^{N-1}(X_n X_{n+1}+Y_n Y_{n+1})+\frac{m}{2}\sum_{n=1}^N (-1)^n Z_n \notag\\
     &
     +\frac{a g^2}{2}\sum_{n=1}^{N-1} (\Ldyn{n}+\Lext{n}(t))^2 \ .
 \end{align}

Our simulations then proceed as follows. We start by finding the ground state $\ket{\Psi_0}$ of the usual massive Schwinger model $H^L(0)$. We then compute the state $\ket{\Psi_t} = \mathcal{T} e^{-i \int_0^t H^L(t') \mathrm{d}t'}\ket{\Psi_0}$ corresponding to the evolution under the time-dependent Hamiltonian $H^L(t)$, with $\mathcal{T}$ being the time-ordering operator. The system is effectively ``quenched" at $\frac{t}{a}=\frac{t_0}{a}=1$, when the external sources are introduced. We then compute different time-dependent expectation values $\langle O \rangle_t \equiv \bra{\Psi_t} O \ket{\Psi_t}$ where $O$ are the operators corresponding to observables of interest.

\textit{Vacuum modification and quantum entanglement between the jets:} 
We measure the local electric charge density, the total electric charge, the scalar fermion density $\langle \bar\psi \psi \rangle$,
the local electric field strength, and the  electric field energy, that are given respectively by
\begin{align}
q_{n,t} \equiv\;&
    \langle \psi^\dagger(a\,n) \psi(a\,n) \rangle_t
    = \frac{\langle Z_n \rangle_t + (-1)^n}{2 a} ,\\
Q_t \equiv\;&
    \int \langle \psi^\dagger(x) \psi(x) \rangle_t \,\mathrm{d}x
    = a\sum_{n=1}^N q_{n,t},\\
\nu_{n,t} \equiv\;&
    \langle \bar\psi(a\,n) \psi(a\,n) \rangle_t
    = \frac{(-1)^n\langle Z_n \rangle_t}{2 a} ,\\
\nu_t \equiv\;&
    \int \langle \bar\psi(x) \psi(x) \rangle_t\,\mathrm{d}x
    = a\sum_{n=1}^N \nu_{n,t},\\
\Pi_{n,t} \equiv\;&
    \langle E(a\, n) \rangle_t
    =  g\,\langle L_n \rangle_t,\\
E_{\text{ele},t} \equiv\;&
    \frac{1}{2}\int \langle E^2(x) \rangle_t\,\mathrm{d}x
    = \frac{a\,g^2}{2}\sum_{n=1}^{N-1} \langle L_n^2 \rangle_t.
\end{align}
The expressions in terms of spin variables are obtained by first staggering the spinors and then using the transformation \eqref{eq:jordan_wigner}, see e.g. appendices of \cite{Kharzeev:2020kgc,Chakraborty:2020uhf} for more details.
We also compute the entanglement entropy between the left- and the right-hand sides of the chain
\begin{equation}
S_{EE}(t) =-\Tr_{A}(\rho_{t,A}\log\rho_{t,A}),
\end{equation}
with $A=\{1,\cdots, N/2\}$ and $B=\{N/2+1,\cdots,N\}$. The operator $\rho_{t,A}=\Tr_B{\rho_t}$ is the partial trace of the time dependent density matrix $\rho_t\equiv\ket{\Psi_{t}}\bra{\Psi_{t}}$ over $B$ [see illustration in Fig.~\ref{fig:1}(left)].

In Fig.~\ref{fig:1}, we show the time evolution of local and global observables respectively, for parameters $N=20$, $m=0.25/a$, and $g=0.5/a$.  In the left panel, we show the full-time evolution of our quantum state. We observe that both the gauge fields and the fermion fields are excited by external sources, and their effects are constrained within the light cone spanned by them. We observe a step-like increase in electric field energy in the right panel. The growth of $\nu_t - \nu_0$ shown in Fig.~\ref{fig:1} indicates the destruction of the (negative) vacuum chiral condensate $\nu_0$ by the propagating jets~\footnote{For the case of static sources, partial destruction of the chiral condensate in Schwinger model was studied in~\cite{Kharzeev:2014xta}.}\footnote{For $N=20$, $m=0.25/a$, and $g=0.5/a$, the value of the vacuum fermion condensate integrated over the lattice length is $\nu_0 = -5.16$. The corresponding value of the vacuum fermion condensate is $\bar{\psi}\psi = -0.258/a$.}. This destruction is due to the pair production from the vacuum that also results in the screening of the electric energy which appears smaller than the contribution from  external sources. We have also performed a comparison to analytical results in the massless fermion case. The results are reported in the Supplementary Material.

Since we can access the entire quantum state, we are able to compute also for the first time the entanglement entropy between the jets. The growth of this entanglement entropy (third panel) results from the pair creation. Lastly, as a consistency check, we also show in the lower panel the total electric charge, which remains zero, as expected.

\textit{Observing quantum entanglement between the jets:} With an eye towards possible experimental studies of quantum entanglement between the produced jets, 
 we measure the two-point correlation of scalar fermion density operators with the vacuum expectation value subtracted,
\begin{align}
    \langle \Delta\nu_{N/2+\ell} \; \Delta\nu_{N/2+1-\ell}\rangle,
    \label{eq:corr}
\end{align}
where $\Delta \nu_n \equiv \nu_n - \langle \nu_n \rangle_\mathrm{vac}$. 

The motivation behind this study is the following. In the bosonization dictionary of the massive Schwinger model, the correlation between the scalar fermion densities translates into the correlation among the boson pairs (and higher order correlations). Therefore we hope that this correlation function may be used to infer information about quantum  entanglement between the pion pairs produced in jet fragmentation. A concrete proposal of an observable correlation between pion pairs produced in jet fragmentation has been put forward in~\cite{Efremov:1995ff}.

\begin{figure}[!hbtp]
    \centering
    \includegraphics[width=0.28\textwidth]{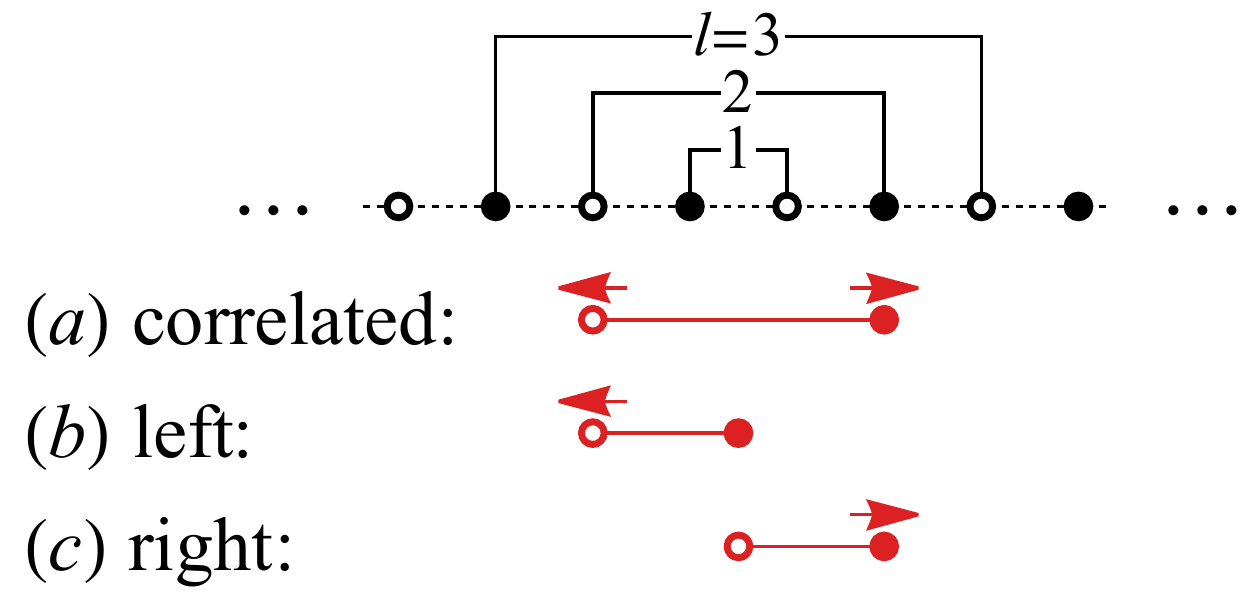}
    \caption{Illustration of correlated and uncorrelated measurements of two point correlation functions. The uncorrelated setup is obtained as an uncorrelated linear superposition of jets created by a single (anti)fermion source moving to the (left)right.}
    \label{fig:3}
\end{figure}

To isolate the effect of entanglement between the jets, we
measure the correlation function for the cases of correlated and uncorrelated sources of fermion-antifermion pairs. Because the entanglement should stem from the correlation between the sources, the case of uncorrelated sources provides the classical baseline for the correlation functions. 
\begin{figure}[!hbtp]
    \centering
    \includegraphics[width=0.45\textwidth]{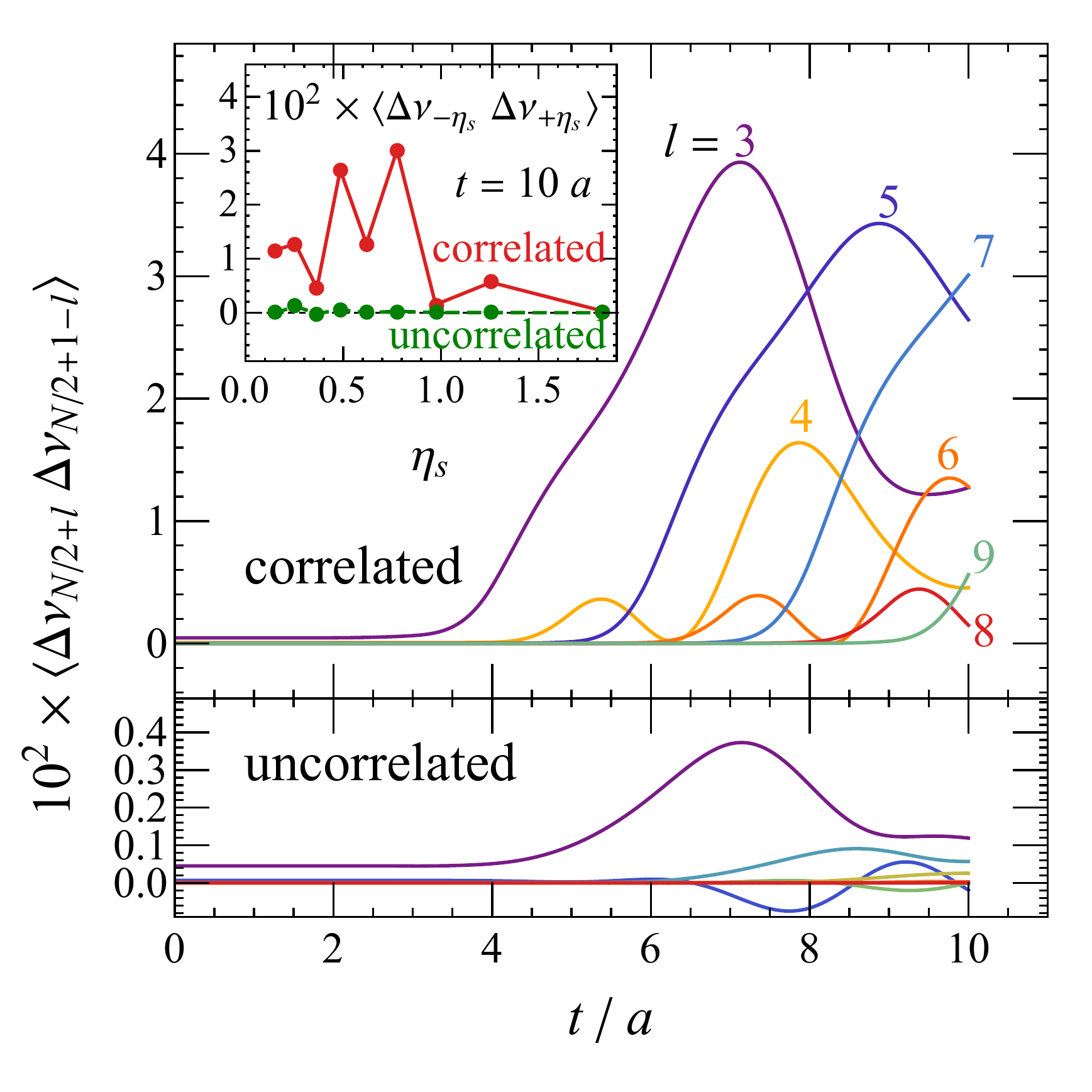}
    \caption{Time evolution of two-point correlation functions with various separations. The upper(lower) panel is for a correlated(uncorrelated) setup. The large difference between the two cases is a signature of quantum entanglement in the produced pairs. (Insert) Spatial-rapidity dependence of the two-point correlation at the end of the evolution.}
    \label{fig:4}
\end{figure}
Our method of preparation of two uncorrelated quantum systems is illustrated in Fig.~\ref{fig:3}~(b, c). In one of these systems, there is only an antifermion source moving to the left while the fermion source sits still at the origin. We denote the quantum state of such a system as $\ket{\psi_\mathrm{L}}$. We then define its counterpart, $\ket{\psi_\mathrm{R}}$, corresponding to the setup of Fig.~\ref{fig:3}(c), with fermion source moving to the right and the antifermion source fixed at the origin. The uncorrelated state is defined as the superposition of left and right state with a random phase, $\ket{\psi_\mathrm{uncorr}} = \frac{1}{\sqrt{2}}\ket{\psi_\mathrm{L}} + \frac{e^{i\varphi}}{\sqrt{2}}\ket{\psi_\mathrm{R}}$, and the expectation value of any observable is obtained by averaging over this random phase, $\langle\langle\psi_\mathrm{uncorr}|O|\psi_\mathrm{uncorr}\rangle\rangle \equiv \int\langle\psi_\mathrm{uncorr}|O|\psi_\mathrm{uncorr}\rangle \frac{\mathrm{d}\varphi}{2\pi}
= \frac{\langle\psi_\mathrm{L}|O|\psi_\mathrm{L}\rangle}{2} + \frac{\langle\psi_\mathrm{R}|O|\psi_\mathrm{R}\rangle}{2}$. The effect of the uncorrelated sources on local charge density and electric field can be found in the Supplemental Materials.

The correlation function~\eqref{eq:corr} is designed to measure the points that are symmetric with respect to the jet production vertex. We measure the two-point correlation function with different separation distances as functions of time, and the results are presented in Fig.~\ref{fig:4}. We find that the correlation functions measured for the correlated state are an order of magnitude greater than those for the uncorrelated state. Note that  it is non-zero in the latter case because of the classical correlation between the particle production in left- and right-moving jets which is similar to the correlation that would be induced by the propagation of sound along the jets' axes. 

Meanwhile, for the quantum correlated state, we observe the propagation of a similar pattern for odd $\ell$'s and similarly for even $\ell$'s, which is driven by the correlated moving sources. After a sufficiently large time, we take a snapshot and present the space dependence of the correlation functions in Fig.~\ref{fig:4}~(insert), where we have converted the site separation to spatial rapidity separation, $\eta_s \equiv \mathrm{arctanh}\frac{x}{t} = \mathrm{arctanh}\frac{(\ell-1/2)a}{t}$. 

One can clearly see a big difference between the strong quantum correlation for the quantum state and the near absence of correlations for the uncorrelated baseline. This difference is especially pronounced for moderate rapidity separations $\Delta \eta_s = 2 \eta_s \leq 2$.

Extrapolating our findings to QCD, we propose to look for quantum entanglement among the pions produced in the fragmentation of the two jets at moderate rapidity separation. An observation of correlations among these pion pairs would constitute a direct signature of entanglement between the jets.
Specifically, it would be interesting to study the quantum correlations between the ``handedness" of the pion pairs produced in the fragmentation of the quark and antiquark jets~\cite{Efremov:1995ff}. Some hints of such correlations had been reported in the data from DELPHI Collaboration~\cite{DELPHICollaboration:1995saf}.

To summarize, we have performed a real-time, non-perturbative study of jet fragmentation using a massive Schwinger model with external sources. Strong distortion of the vacuum chiral condensate by the propagating jets has been observed. We have also found strong quantum entanglement between the fragmenting jets for rapidity separation $\Delta \eta \leq 2$. Our work paves the way for quantum simulations of jet fragmentation using quantum hardware; we plan to address this problem in the near future.

\section*{Acknowledgement}
We thank Jo\~ao Barata, Fangcheng He, Yuta Kikuchi, Semeon Valgushev, Tzu-Chieh Wei, and Ismail Zahed for useful discussions and communications.
This work was supported by the U.S. Department of Energy, Office of Science, National Quantum Information Science Research Centers, Co-design Center for Quantum Advantage (C2QA) under Contract No.DE-SC0012704 (AF, KI, DK, VK), the U.S. Department of Energy, Office of Science, Office of Nuclear Physics, Grants Nos. DE-FG88ER41450 (DF, DK, SS) and DE-SC0012704 (AF, DK, KY), and Tsinghua University under grant no. 53330500923 (SS).
This research used resources of the National Energy Research Scientific Computing Center, a DOE Office of Science User Facility supported by the Office of Science of the U.S. Department of Energy under Contract No. DE-AC02-05CH11231 using NERSC award NERSC DDR-ERCAP0022229.

\clearpage
\begin{widetext}
\begin{center}
\textbf{\large Supplementary Material}
\end{center}
\input{appendix}
\end{widetext}

\bibliographystyle{utphys}
\bibliography{main.bib}

\end{document}

%% file: appendix.tex
\section{Lattice size dependence and boundary effects}
In the main text, we study the evolution of the Schwinger model Hamiltonian in the presence of external charges moving on the light-cone. 
In this supplemental material, we show that despite the relatively modest lattice sizes, the volume dependence and effect of open-boundary conditions are well under control for the quantities and set of parameters we studied.

\begin{figure*}[!htp]
    \centering
    \includegraphics[width=0.31\textwidth]{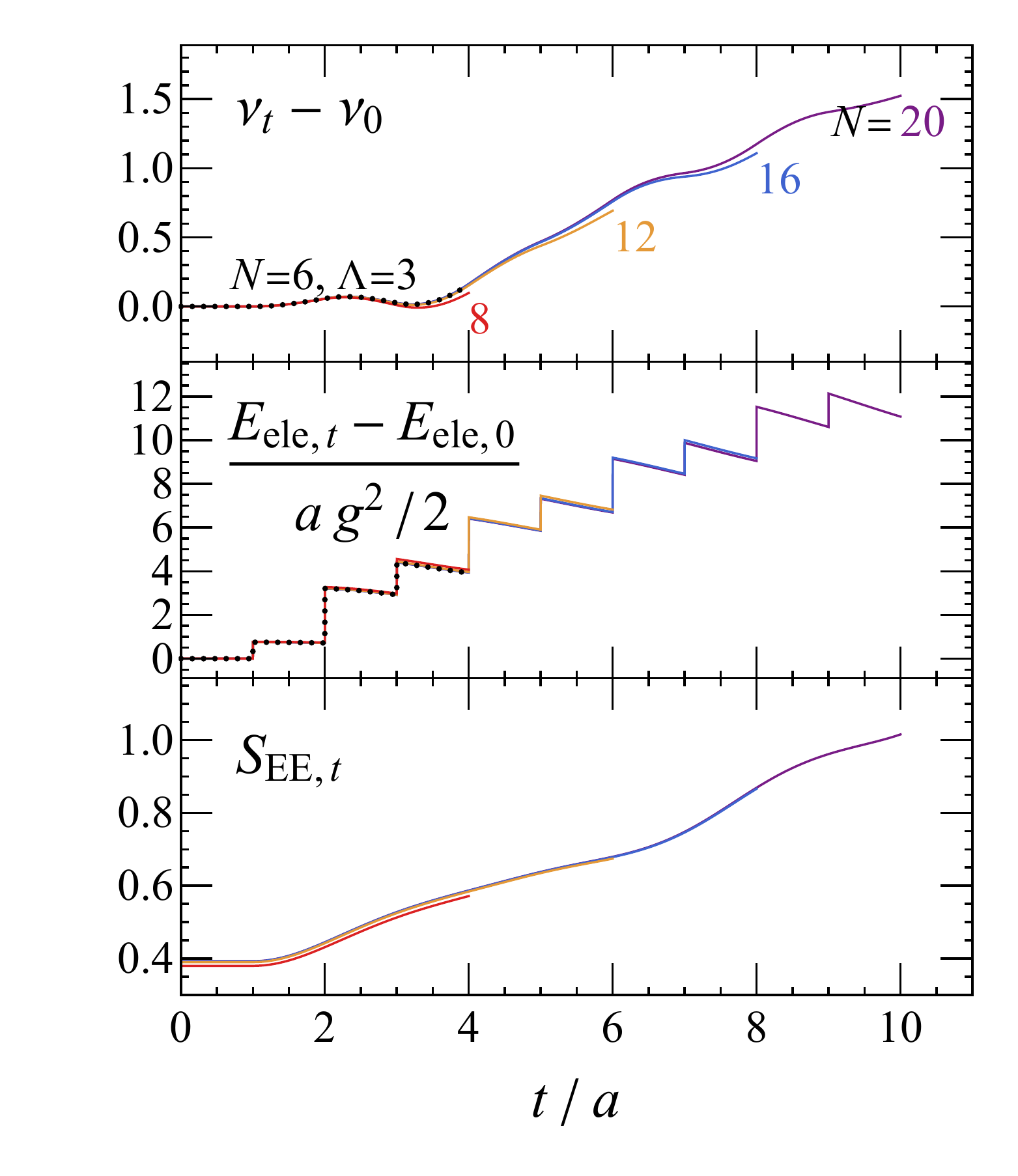}
    \includegraphics[width=0.31\textwidth]{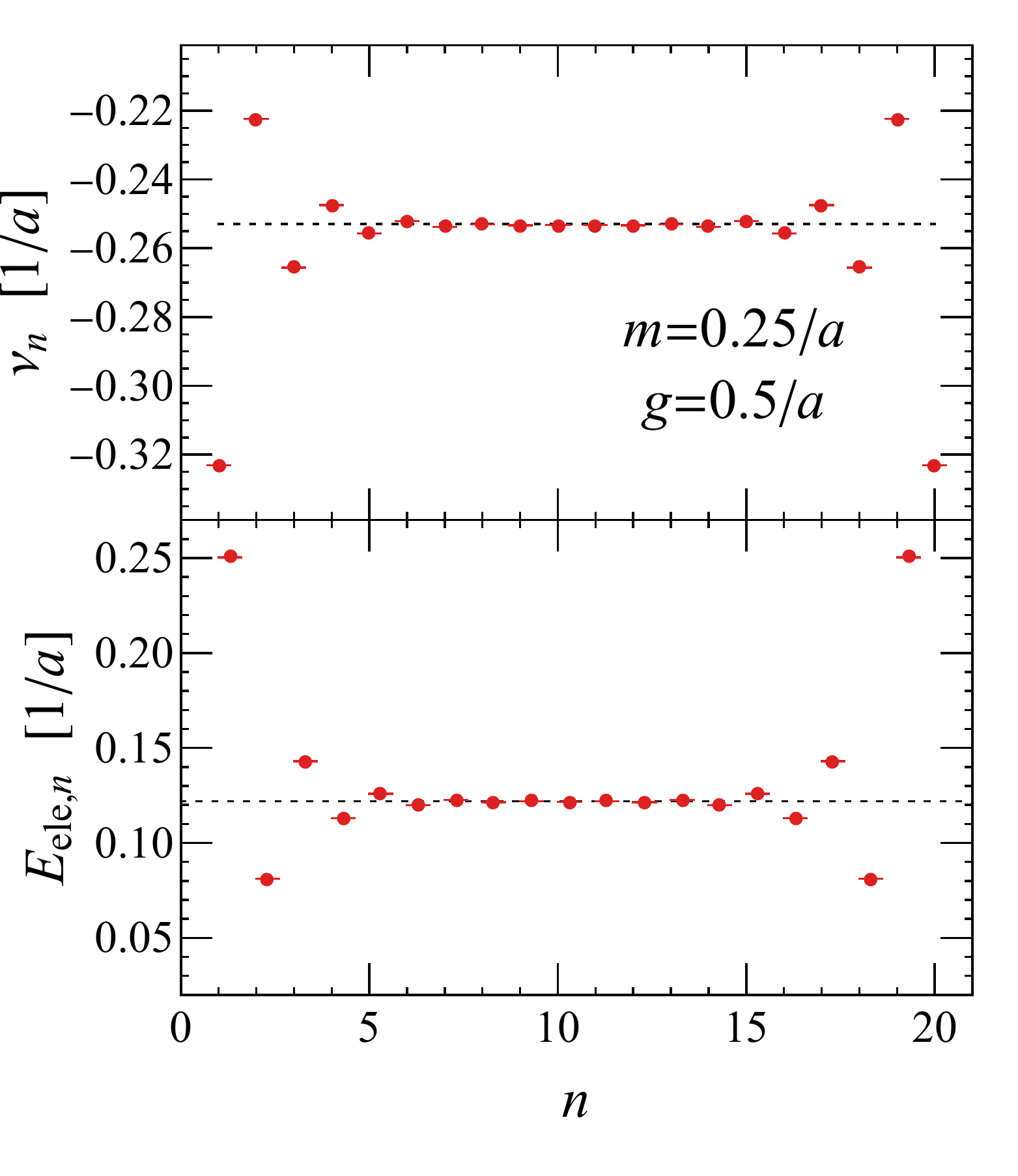}
    \includegraphics[width=0.31\textwidth]{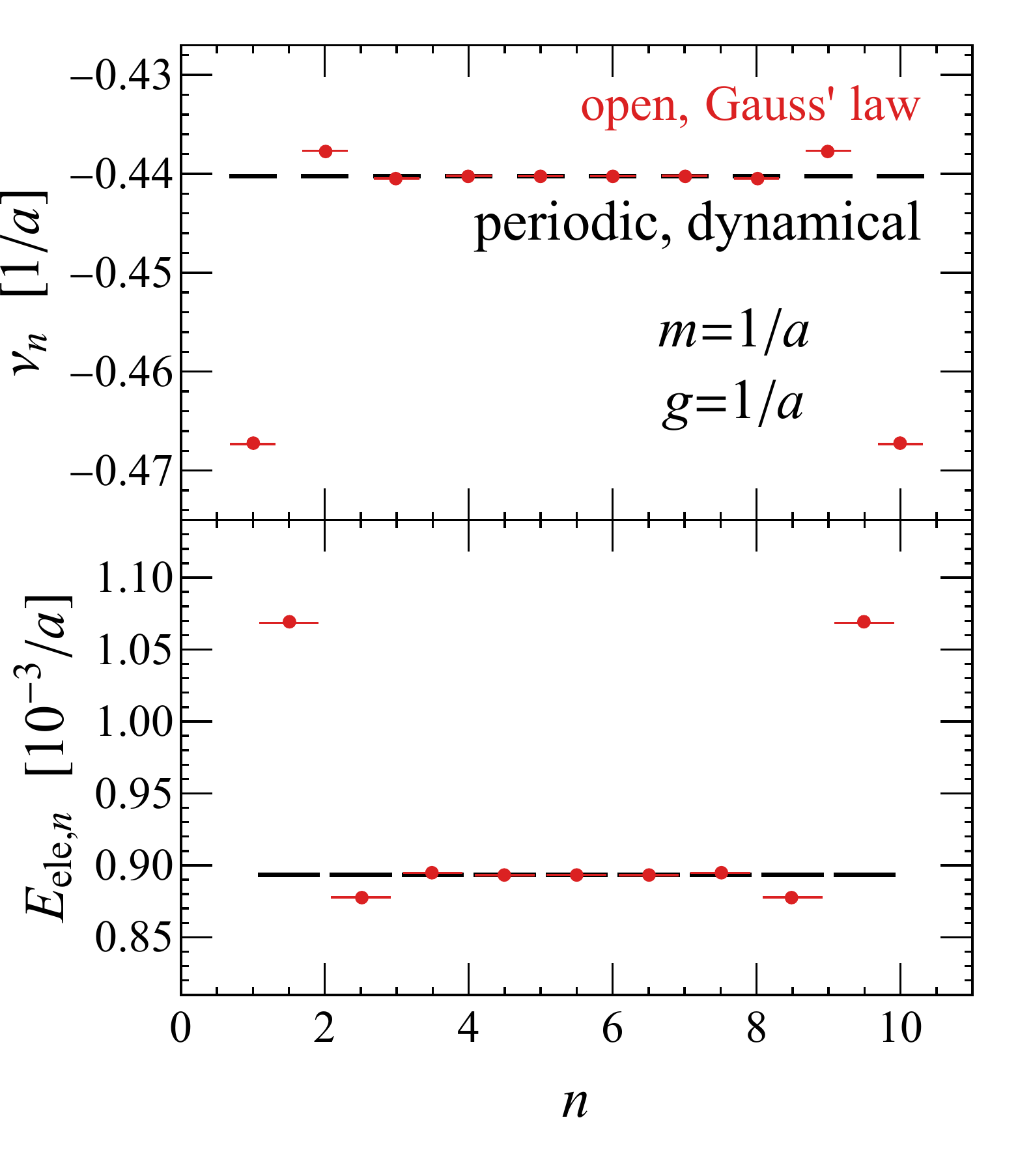}
    \caption{
    (Left) Time evolution of total electric field energy, mass creation, and entanglement entropy for periodic boundary condition with dynamical gauge field with $N=6$ and $\Lambda=3$ (black dotted) versus open boundary condition with gauge field fixed by the Gauss' law with lattice size from $8$(red) to $20$ (purple).
    (Middle) Comparison of local electric field energy and chiral condensate.
    Black dotted lines are determined by the bulk values.
    In both left and middle panels, parameters are set to be $N=20$, $m=0.25/a$, and $g=0.5/a$.
    (Right) Same as middle but with parameters with parameters $N=10$, $m=1/a$, and $g=1/a$.
    Red dots correspond to open boundary condition with gauge field fixed by the Gauss' law, whereas black lines are for periodic boundary condition with dynamical gauge field.
    }
    \label{fig:comparison}
\end{figure*}

In the left-hand side of Fig.~\ref{fig:comparison}, plain colored lines show the time evolution of the chiral condensate, electric field energy and entanglement entropy for different lattice sizes. The maximal time until which a simulation is meaningful is set by half the lattice site plus one unit of time, as after this the point sources exit the system. As illustrated by the agreement of the different curves, finite size effects are minimal.

We also assess the effect of using open-boundary conditions.  We expect that the introduction of a physical boundary to have the same effect as the introduction of a defect. Excitations localize on the boundary and affect the system in a ``boundary zone" of order the correlation length of the system, see for instance \cite{Bruno2016The}.  We can see in the middle panel of  Fig.~\ref{fig:comparison} that this is indeed what happens. We show in the upper(lower) panel the value of the chiral condensate(electric energy density) as a function of lattice sites in the ground state. In both cases, we can clearly observe a boundary zone extending over approximately 4-5 lattice sites. It also matches the naive estimate of the correlation length  $\xi\sim \frac{1}{m} = 4 a$. 

To further crosscheck our results, we also decided to implement simulations with periodic-boundary conditions, $\chi_{N+1}=\chi_{1}$ and $\chi_{N+1}^\dagger=\chi_{1}^\dagger$,
and to keep the gauge field as independent operators. The Hamiltonian reads
\begin{align}
\begin{aligned}
     H_\text{PBC}=&\frac{1}{8a}\sum_{n=1}^{N}\Big[(U_n+U^\dagger_n)\otimes(X_n X_{n+1}+Y_n Y_{n+1})
     +i(U_n-U^\dagger_n)\otimes(X_nY_{n+1}-Y_nX_{n+1})\Big]\\
     &+\frac{m}{2}\sum_{n=1}^N(-1)^n Z_n+\frac{a\ g^2}{2}\sum_{n=1}^{N} L^2_n
     +\frac{1}{g}\sum_{n=1}^{N} j^1_{ext}(a\, n)\phi_n \,,
\end{aligned}
\end{align}
where $X_{N+1} \equiv (-1)^{\frac{N}{2}} X_1 \prod_{m=2}^{N-1} Z_m$, and likewise for $Y_{N+1}$. We implement the electric-field operator and the link operator as
\begin{align}
    L_n&=\sum_{\epsilon=-\Lambda}^{\Lambda}\epsilon\ket{\epsilon}_n\bra{\epsilon}_n \,,\\
    U_n&=
    \ket{\Lambda}_n \bra{-\Lambda}_n +
    \sum_{\epsilon=-\Lambda}^{\Lambda-1}\ket{\epsilon}_n\bra{\epsilon+1}_n \,,
\end{align}
where $\Lambda$ is a cutoff~\cite{Shaw:2020udc}, the eigenbasis $\ket{\epsilon}_n$ of electric field operator $L_n$. 

The size of the discrete Hilbert space for a truncation  $\Lambda$ is $(2\Lambda+1)^N \; 2^N$, namely it is $(2\Lambda+1)^N$ times larger than in the case of open boundary conditions after integrating out the gauge fields through Gauss law. This also means that only smaller lattices can be simulated in this set-up. 

We show results of the chiral condensate and electric field energy for $N=6$ and $\Lambda=3$ as black dotted lines in the left-hand side of Fig.~\ref{fig:comparison}. No deviations from the open-boundary conditions can be seen. 

We also investigated the space-dependence of observables. In particular, we expect the bulk value of the open-boundary conditions to equal the periodic boundary condition average. Unfortunately, we could not directly verify this for the parameters used in the main text as the lattice size required are not achievable not integrating out gauge fields. As an alternative, we verified it for a larger mass and larger coupling $ma = ga = 1$ such that the boundary zone is smaller. The results are shown in the right-hand side panel of Fig.~\ref{fig:comparison}. Again, the two lattice sites affected by the boundary is in agreement with  naive expectations. And as expected, the bulk value of the open-boundary system matches the value of the periodic one.

\clearpage
\section{Uncorrelated Sources}
\begin{figure}[!hbtp]\centering
    \includegraphics[width=0.49\textwidth]{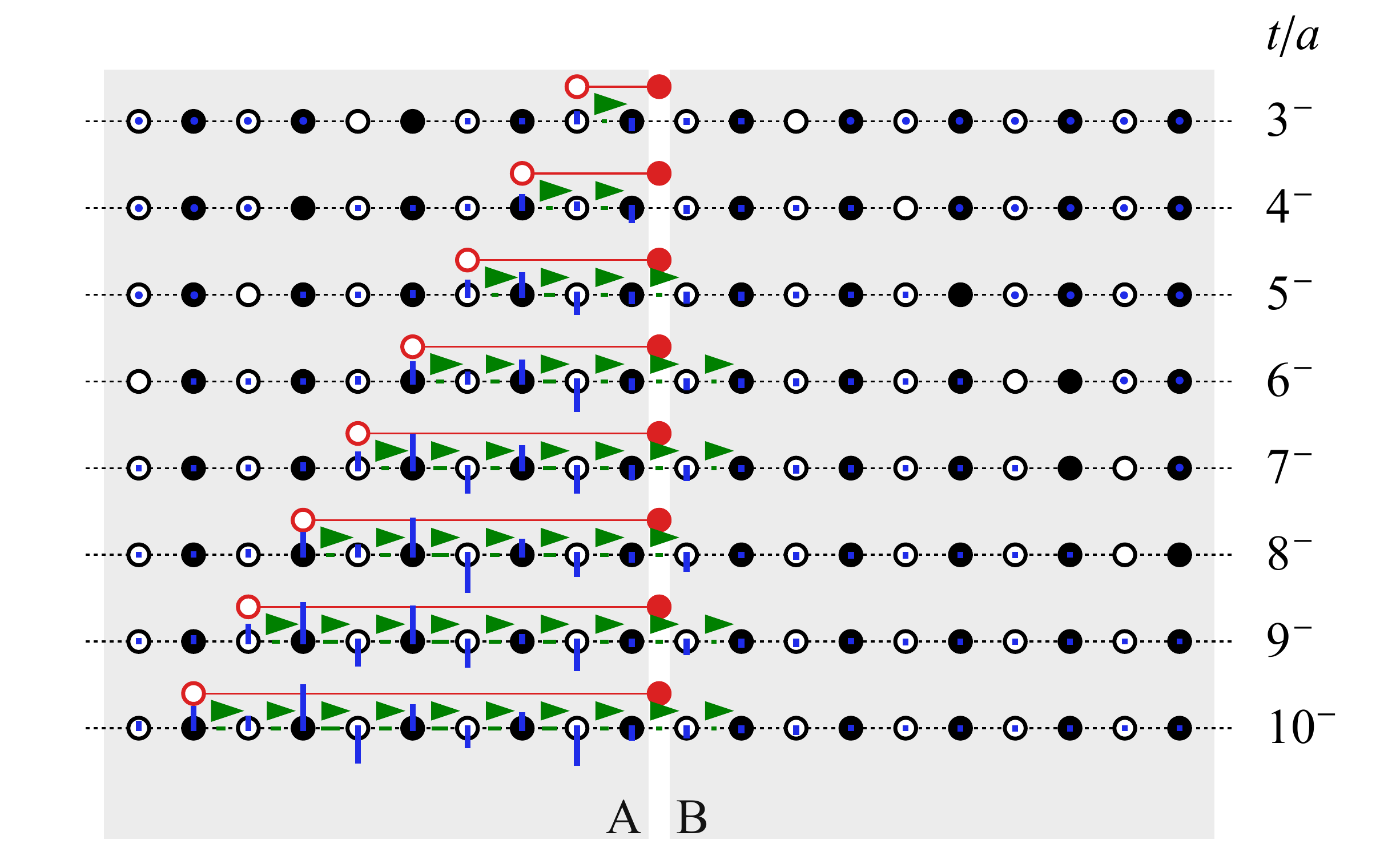}
    \includegraphics[width=0.49\textwidth]{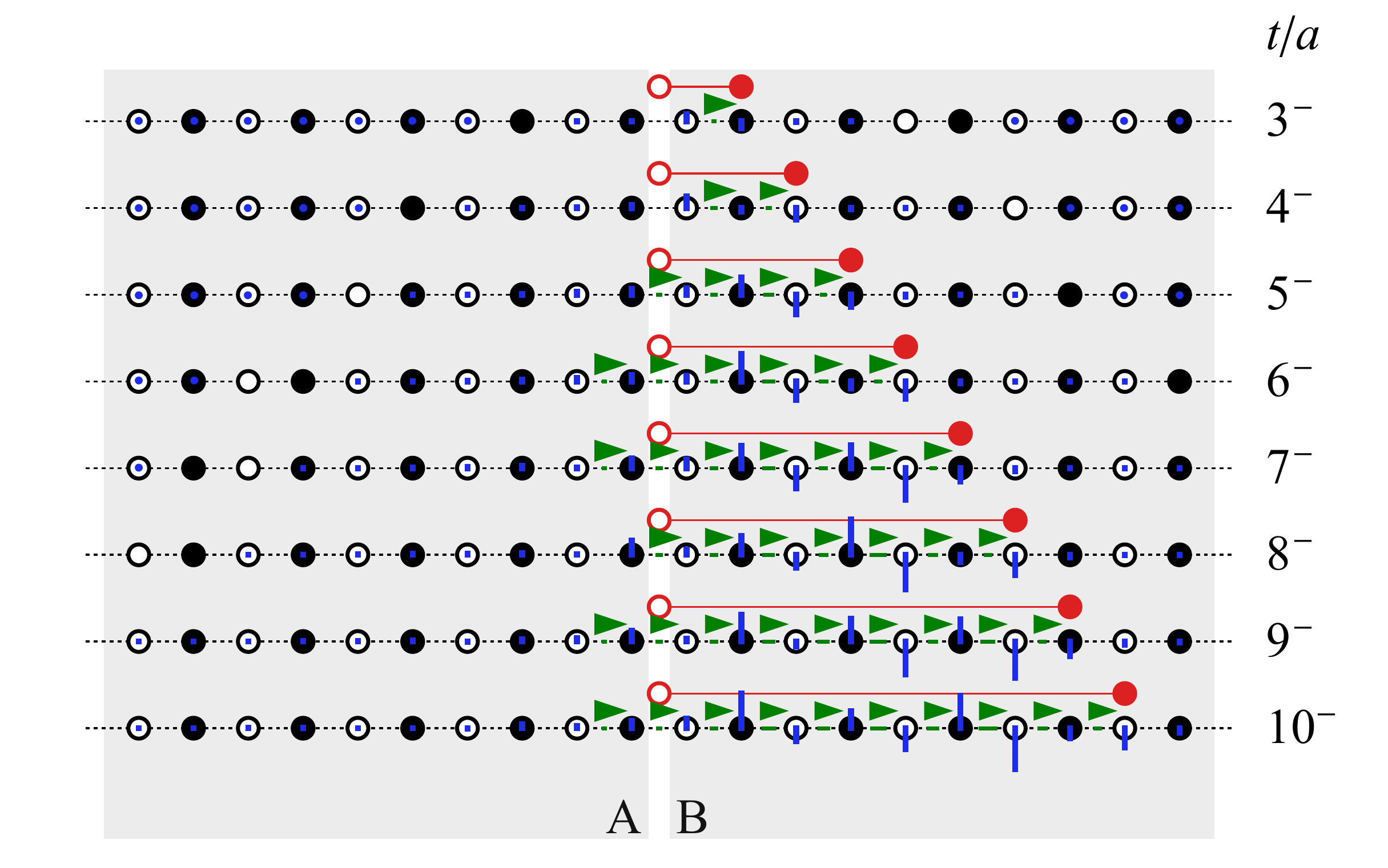}\\
    \includegraphics[width=0.49\textwidth]{fig1}\\
    \caption{Same as Fig.~1 (left) of the main text but for left-moving(upper left) and right-moving(upper right) uncorrelated sources. Fig.~1 (left) is repeated as the lower panel, for convenient comparison.
    \label{fig:uncorrelated}}
\end{figure}

In the main text we prepared uncorrelated left-moving and right moving external sources to provide the baseline of correlation functions without quantum entanglement. The response of the system is shown in the upper panels of Fig.~\ref{fig:uncorrelated}. 
For better comparison, we also include the response to the correlated source, i.e. Fig.~1~(left) of the main text. 
We observe that local charge densities and electric fields in the case with left-moving source [section ``A'' in Fig.~\ref{fig:uncorrelated}~(upper-left)] are very similar to those of the left half plane under correlated sources [section ``A'' in Fig.~\ref{fig:uncorrelated}~(lower)].
Corresponding similarity is also observed in the right-moving counter parts, i.e., sections ``B'' of Fig.~\ref{fig:uncorrelated}~(upper-left) and Fig.~\ref{fig:uncorrelated}~(lower).
Thus, the uncorrelated state preparation provides almost identical results for $\langle \nu_{n} \rangle$, and one could therefore expect reliable estimation for the classical baseline of the correlation functions.

\clearpage
\section{Comparison to analytic solution at zero fermion mass}

\begin{figure}[!htbp]
\centering
\includegraphics[width=\textwidth]{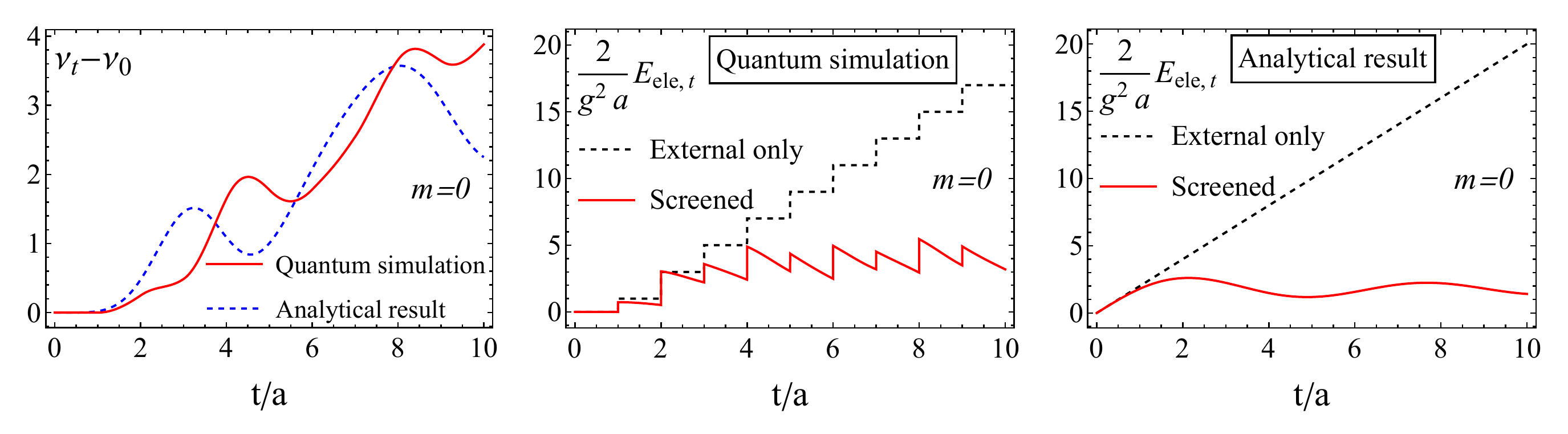}
\caption{(Left) Comparison between the deviation of chiral condensate from its vacuum value in the quantum simulation and analytical solution. (Middle) Screening of the electric field produced by the external sources in the quantum simulation. (Right) Screening of the electric field produced by the external sources in the analytical solution. In all panels the parameters are $g=1/a, m=0$. Quantum simulation is performed for $N=20$.}
\label{fig:massless}
\end{figure}
As noted in the main text, it was first suggested to model jets with two external charges in the context of massless Schwinger model in \cite{Casher:1974vf} with the subsequent developments in \cite{Loshaj:2011jx, Kharzeev:2012re}. Since massless Schwinger model is exactly solvable, some of the quantities studied in our numerical analysis can be computed analytically there. Notably, we can do so with the chiral condensate and the electric field energy. The purpose of this Appendix is to provide a comparison between those analytical results and the ones reported in the main text.

The simplest formalism which allows to solve the massless Schwinger model exactly is bosonization. According to the bosonization dictionary massless Schwinger model is equivalent to a free massive scalar field. In the presence of an external source the Lagrangian of the theory in the continuum is (\cite{Coleman:1975pw})
\beq \label{eq:lagrangian}
\mathcal{L} = \frac{1}{2}(\partial_\mu \phi)^2 - \frac{1}{2} m_b^2 (\phi + \phi_{ext})^2,
\eeq
where $m_b = \dfrac{g}{\sqrt{\pi}}$ is the boson mass expressed through the coupling constant of the Schwinger model. The external source field is obtained by the correspondence between fermionic current and gradients of the scalar field:
\beq
j^\mu_{ext} = \frac{1}{\sqrt{\pi}} \epsilon^{\mu\nu}\partial_\nu\phi_{ext},
\eeq
which for the external current corresponding to two opposite charges propagating along the light cone, eq.(\ref{eq:jext}), leads to
\beq \label{eq:massless_source}
\phi_{ext}(t,x) = -\sqrt{\pi} \theta(t^2 - x^2).
\eeq
Since the bosonized model is free, we can extract all the necessary information from a classical solution to the equation of motion in the presence of the external source:

\beq
(\Box + m_b^2) \phi(t,x) = - m_b^2 \phi_{ext}(t,x).
\eeq

The solution to this equation which has the correct boundary conditions is (note that to be consistent with the current normalization we restored a factor of $\sqrt{\pi}$ compared to \cite{Kharzeev:2012re})
\beq \label{eq:massless_analytic}
\phi(t,x) = \sqrt{\pi}\theta(t^2-x^2) \left[1-J_0\left(m_b\sqrt{t^2-x^2}\right)\right].
\eeq

Bosonization prescribes the following expressions for the chiral condensate and the electric field in terms of $\phi$:
\begin{align}
\begin{aligned}
\bar{\psi}\psi(x) = - c\, m_b\cos[2\sqrt{\pi}\phi(x)], \\
E(x) = - m_b[\phi(x)+\phi_{ext}(x)],
\end{aligned}
\end{align}
where $c = \dfrac{e^\gamma}{2\pi}$ and $\gamma$ is the Euler constant. We plug the expressions (\ref{eq:massless_analytic}) and (\ref{eq:massless_source}) into the integrals over space of the chiral condensate deviation from the vacuum and electric field energy, normalized according to the main text:
\begin{align}
\begin{aligned}
\nu_t - \nu_0 = \int_{-\infty}^\infty dx [
\langle\bar{\psi}(x)\psi(x)\rangle_t-\langle\bar{\psi}(x)\psi(x)\rangle_{vac}] =c\, m_b \int_{-t}^t dx \left[1-\cos\left(2\pi\left[1-J_0\left(m_b\sqrt{t^2-x^2}\right)\right]\right)\right]  \\
\frac{2}{g^2 a} E_{ele, t} = \frac{1}{g^2 a}\int_{-\infty}^\infty dx \langle E^2(x)\rangle_t = \frac{1}{a}\int_{-t}^t dx \left[J_0\left(m_b\sqrt{t^2-x^2}\right)\right]^2
\end{aligned}
\end{align}

These integrals are evaluated numerically and displayed on Fig.~\ref{fig:massless} as functions of $t$. To make contact with the lattice model we choose to measure $x$ and $t$ in terms of $a$ which is simply set to 1. The coupling constant is set to $g=\dfrac{1}{a} = 1$.

For comparison, we perform a quantum simulation in the same way as described in the main text for parameters $m=0, g=\dfrac{1}{a}, N=20$. The presence of a gap, evident from the bosonic formulation, leads to a finite correlation length, rendering simulations in a finite volume possible. On the left panel of Fig.~\ref{fig:massless} we show the growth of the chiral condensate. We see that both curves are in qualitative agreement. The quantitative differences arise from discretization effects at finite $a$ and are expected to disappear in the continuum limit. Unfortunately, our current approach does not allow us to take the continuum limit, as it would require working with larger volumes. 

On the middle and right panels of Fig.~\ref{fig:massless} we compare the screening of the external electric field in the quantum simulation and in the analytical solution. The staircase structure of the numerical simulation result is again an effect of the discretization. The step size is equal to the lattice spacing and goes to zero in the continuum limit. Except for this effect, numerical results are in good agreement with the continuum analytical expression. Notably, in both approaches the electric field energy stays roughly constant after a short initial time. This is a signature of a very strong screening that is expected for massless fermions where pairs are easily produced. A plausible interpretation is that apart from small transitional regions nearby the external charges electric field is completely screened farther away from the charges.